\documentclass[a4paper,11pt]{article}
\pdfoutput=1
\usepackage{jcappub}
\usepackage{slashed}
\usepackage{mathtools}

\newcommand{\gs}{g_\star}
\newcommand{\gss}{g_{\star s}}
\newcommand{\Trh}{T_\text{rh}}
\newcommand{\arh}{a_\text{rh}}
\newcommand{\mdm}{m_\text{dm}}
\newcommand{\Tmax}{T_\text{max}}
\newcommand{\Gp}{\Gamma_\phi}
\newcommand{\rp}{\rho_\phi}
\newcommand{\rR}{\rho_R}
\newcommand{\np}{n_\phi}
\newcommand{\ndm}{n_\text{dm}}
\newcommand{\nnp}{n_\text{np}}
\newcommand{\mphi}{m_\phi}
\newcommand{\br}{\mathcal{B}}

\title{Rescuing Gravitational-Reheating\\in Chaotic Inflation}
\author[a,b]{Basabendu Barman,}
\author[c]{Nicolás Bernal}
\author[d]{and Javier Rubio}
\affiliation[a]{Institute of Theoretical Physics, Faculty of Physics, University of Warsaw\\ ul. Pasteura 5, 02-093 Warsaw, Poland}
\affiliation[b]{Department of Physics, School of Engineering and Sciences,
SRM University-AP, Amaravati 522240, India}
\affiliation[c]{New York University Abu Dhabi\\
PO Box 129188, Saadiyat Island, Abu Dhabi, United Arab Emirates}
\affiliation[d]{Departamento de Física Teórica and\\
Instituto de Física de Partículas y del Cosmos (IPARCOS-UCM)\\
Universidad Complutense de Madrid, 28040 Madrid, Spain}
\emailAdd{basabendu.b@srmap.edu.in}
\emailAdd{nicolas.bernal@nyu.edu}
\emailAdd{javier.rubio@ucm.es}
\abstract{
We show, within the single-field inflationary paradigm, that a linear non-minimal interaction $\xi\,M_P\,\phi\,R$ between the inflaton field $\phi$ and the Ricci scalar $R$ can result in successful inflation that concludes with an efficient heating of the Universe via perturbative decays of the inflaton, aided entirely by gravity. Considering the inflaton field to oscillate in a quadratic potential, we find that $\mathcal{O}(10^{-1}) \lesssim \xi \lesssim \mathcal{O}(10^2)$ is required to satisfy the observational bounds from Cosmic Microwave Background (CMB) and Big Bang Nucleosynthesis (BBN). Interestingly, the upper bound on the non-minimal coupling guarantees a tensor-to-scalar ratio $r \gtrsim 10^{-4}$, within the range of current and future planned experiments. We also discuss implications of dark matter production, along with the potential generation of the matter-antimatter asymmetry resulting from inflaton decay, through the same gravity portal.  
}
\begin{document}
\begin{flushright}
    IPARCOS-UCM-23-118
\end{flushright}
\maketitle

%%%%%%%%%%%%%%%%%%%%%%%%%%%%%%%%%%%%%
\section{Introduction}
%%%%%%%%%%%%%%%%%%%%%%%%%%%%%%%%%%%%%
Inflation serves as a well-established framework that harmoniously aligns with our empirical observations, offering elegant solutions to the puzzles within the hot Big Bang model~\cite{Guth:1980zm, Linde:1981mu}. In its simplest avatar, the inflationary stage is driven by a slowly rolling scalar field whose energy density dominates the Universe at some early epoch and is eventually converted into a radiation bath, (re)heating the Universe and signaling the onset of radiation domination~\cite{Bassett:2005xm, Allahverdi:2010xz, Amin:2014eta, Lozanov:2019jxc}. Depending on the model under consideration, this relocation of energy can take place via perturbative decays~\cite{Albrecht:1982mp, Dolgov:1982th, Abbott:1982hn} or involve highly nonlinear and non-perturbative effects such as parametric resonance~\cite{Traschen:1990sw, Kofman:1994rk, Kofman:1997yn, Greene:1997fu}, tachyonic instabilities~\cite{Felder:2000hj, Felder:2001kt, Dufaux:2006ee, Opferkuch:2019zbd, Bettoni:2021zhq, Laverda:2023uqv}, oscillon formation~\cite{Amin:2010dc, Amin:2011hj, Lozanov:2017hjm, Lozanov:2019ylm, Piani:2023aof} and turbulent energy cascades~\cite{Micha:2002ey, Micha:2004bv}, in isolation or co-existence~\cite{Repond:2016sol, Fan:2021otj}. 

As has recently been pointed out~\cite{Haque:2022kez}, an irreducible Planck suppressed coupling between all matter fields and gravity can lead to gravity-mediated heating, which has been named as ``gravitational reheating'' scenario. As shown in Refs.~\cite{Haque:2022kez, Co:2022bgh, Barman:2022qgt, Clery:2023mjo}, for an inflaton $\phi$ oscillating in a monomial potential $V(\phi)\propto\phi^k$, the minimal gravitational heating scenario, where a pair of inflaton condensate excitations scatters via massless gravitons into standard model (SM) particles (like the Higgs boson), requires $k>9$. Interestingly enough, this bound can be relaxed to $k>4$ if one introduces a non-minimal {\it quadratic} coupling between gravity and the scalars of the theory. However, gravity-mediated gravitational heating through 2-to-2 scattering remains still not viable if the inflaton oscillates at the minimum of a quartic $(k=4)$ or a quadratic $(k=2)$ potential. 

In this work, we will explore a scenario where successful inflation, together with heating, can be achieved through the non-minimal {\it linear} interaction $\xi\, M_P\, \phi\, R$ between the inflaton field $\phi$ and the Ricci scalar $R$, where $M_P$ is the reduced Planck mass and $\xi$ the non-minimal coupling. In particular, we will assume the inflaton field to oscillate in a simple quadratic potential, showing explicitly that this setting can give rise to an adequate number of $e$-folds of inflation and to the onset of radiation domination prior to Big Bang Nucleosynthesis (BBN)~\cite{Sarkar:1995dd, Kawasaki:2000en, Hannestad:2004px, DeBernardis:2008zz, deSalas:2015glj, Hasegawa:2019jsa}, such that the standard cosmological lore is not hampered. Interestingly, both the inflationary and heating dynamics are governed by a single free parameter $\xi$.

Moreover, the inadequacy of the SM in offering a viable dark matter (DM) candidate necessarily calls for beyond the SM fields. Measurements of anisotropies in the cosmic microwave background radiation (CMB) provides the most precise measurement of the DM relic density, usually expressed as $\Omega_\text{DM}h^2 \simeq 0.12$~\cite{Planck:2018vyg}, which the candidate for DM must satisfy. Now, irreducible gravitational interaction can lead to inevitable DM production (or production of {\it any} particle in general), commonly known as ``gravitational production'' of DM.\footnote{Purely gravitational production of particles beyond the SM can also emerge from Hawking radiation of evaporating primordial black holes; see, for example, Refs.~\cite{Green:1999yh, Khlopov:2004tn, Dai:2009hx, Fujita:2014hha, Allahverdi:2017sks, Lennon:2017tqq, Morrison:2018xla, Hooper:2019gtx, Chaudhuri:2020wjo, Masina:2020xhk, Baldes:2020nuv, Gondolo:2020uqv, Bernal:2020kse, Bernal:2020ili, Bernal:2020bjf, Bernal:2021akf, Cheek:2021odj, Cheek:2021cfe, Bernal:2021yyb, Bernal:2021bbv, Barman:2021ost, Barman:2022gjo, Barman:2022pdo, Bernal:2022oha, Cheek:2022dbx, Mazde:2022sdx, Cheek:2022mmy}.} For instance, the production of purely gravitational DM due to expansion of the Universe (which is the conventional gravitational production) has been extensively discussed in Refs.~\cite{Tang:2016vch, Ema:2018ucl, Hashiba:2018tbu}, through the $s$-channel exchange of massless gravitons in, e.g., Refs.~\cite{Garny:2015sjg, Garny:2017kha, Tang:2017hvq, Bernal:2018qlk, Barman:2021ugy, Clery:2021bwz, Mambrini:2021zpp, Clery:2022wib, Barman:2022qgt, Ahmed:2022tfm, Haque:2023yra, Kaneta:2023uwi}, while DM is sourced from the decay of the inflaton in Refs.~\cite{Ellis:2015jpg, Kaneta:2019zgw, Bernal:2021qrl,Barman:2022tzk}. Being a purely gravitational process, the corresponding DM yield is Planck suppressed and can only be dominant at high temperatures.

Finally, the observed excess of baryons over antibaryons in the Universe is quantified in terms of the baryon-to-photon ratio $\eta_B \simeq 6.2 \times 10^{-10}$~\cite{Planck:2018vyg}, based on CMB measurements, which also agrees well with the BBN estimates~\cite{ParticleDataGroup:2020ssz}. 
Although it has all the necessary components, the SM does not satisfy the Sakharov conditions~\cite{Sakharov:1967dj} necessary to generate the adequate asymmetry, demanding physics beyond the SM.
An intriguing possibility to achieve baryogenesis (that is, the dynamical generation of the baryon asymmetry of the Universe (BAU)) is known as leptogenesis~\cite{Fukugita:1986hr} where, instead of explicitly creating a baryon asymmetry, first a lepton asymmetry is produced that subsequently converts into baryon asymmetry by the $(B+L)$-violating electroweak sphaleron transitions~\cite{Kuzmin:1985mm}.

In the present context, both the DM and the observed BAU (along with the right active neutrino mass) can be generated from decays of the inflaton,\footnote{Such decay can also give rise to high-frequency gravitational waves via graviton bremsstrahlung as discussed in Refs.~\cite{Nakayama:2018ptw, Huang:2019lgd, Ghoshal:2022kqp, Barman:2023ymn, Barman:2023rpg, Bernal:2023wus}.} once beyond the SM fields are introduced.\footnote{Gravitational production of DM, along with the BAU, has also been addressed in Refs.~\cite{Bernal:2021kaj, Barman:2022qgt}.} This also falls under the category of gravitational production, as in the absence of the nonminimal coupling to gravity, such a production channel ceases to exist.
We emphasize that such linear non-minimal coupling has not been widely discussed in the literature in the context of inflation and heating.\footnote{In context of DM decay this has been discussed in Refs.~\cite{Cata:2016dsg, Cata:2016epa, Bezrukov:2020wnl}.}

This paper is organized as follows. In Section~\ref{sec:inf} we discuss our model construction together with the inflationary framework. In Section~\ref{sec:heating} we elaborate on the heating scenario and investigate the production of DM and baryon asymmetry. Finally, we conclude in Section~\ref{sec:concl}. 

%%%%%%%%%%%%%%%%%%%%%%%%%%%%%%%%%%%%%
\section{Inflation with a Linear Non-minimal Coupling} 
\label{sec:inf}
%%%%%%%%%%%%%%%%%%%%%%%%%%%%%%%%%%%%%
Let us consider the following action for a secluded weakly self-interacting $\mathbb{Z}_2$ symmetric\footnote{In particular, we require the strength $\lambda$ of quartic self-interaction $\lambda\,\phi^4$ to be $\lambda\ll m_\phi^2/\phi_*^2$, with $\phi_*\sim {\cal O}(M_P)$ the value of the field associated the horizon crossing of the pivot scale. The same kind of fine tuning applies also to higher-order interactions $\sum_n \alpha_n {\tilde{\phi}}^n/M_P^{n-4}$ with $n>4$.} inflaton field $\tilde\phi$ of mass $m_\phi$, non-minimally coupled to gravity and without tree-level interactions to the SM states,
\begin{equation}\label{Jframe}
    S_\phi = \int d^4x\, \sqrt{-\tilde g} \left[\frac12\, M_P^2\, F(\tilde \phi)\, \tilde g^{\mu\nu}\,\widetilde {R}_{\mu\nu}(\widetilde\Gamma)  + \frac12\, \tilde g^{\mu\nu}\, \partial_\mu \tilde\phi\, \partial_\nu\tilde\phi -  \frac12\,m_\phi^2\,\tilde\phi^2  
    \right].
\end{equation}
Here we have adopted a mostly-plus convention for the metric $\tilde g_{\mu\nu}$, $M_P \simeq 2.44\times 10^{18}$~GeV stands for the reduced Planck mass, $F(\tilde \phi)$ is a general function of $\vert \tilde \phi\vert/M_P$ parametrizing all potential non-linearities in the scalar sector and admitting a expansion around unity\footnote{The modulus could be associated, for instance, with a non-linear interaction $\sqrt{\tilde \phi^2+\epsilon}$, with small $\epsilon$.} and
\begin{equation}\label{eq:connect}
    \widetilde R_{\mu\nu}= \partial_\sigma  \widetilde\Gamma_{\ \mu\nu}^\sigma -\partial_\mu \widetilde\Gamma^\sigma_{\ \sigma\nu} + \widetilde\Gamma^\rho_{\ \mu\nu} \widetilde\Gamma^\sigma_{\ \sigma\rho}- \widetilde\Gamma^\rho_{\ \sigma\nu} \widetilde\Gamma^\sigma_{\ \mu\rho}
\end{equation}
denotes the Ricci tensor constructed out of a connection $\widetilde\Gamma^\rho{}_{\mu\nu}$, to be specified in what follows.
Note that, for negligible non-minimal couplings $F(\tilde\phi)\simeq 1$, this action reduces to a particularly simple chaotic scenario, the seminal quadratic inflation model, where the only free parameter $m_\phi$ is completely determined by the measured amplitude of the primordial power spectrum of density fluctuations. However, this simplified scenario scenario is in conflict with the combined Planck and BICEP2/Keck bound on the tensor-to-scalar ratio, namely $r<0.032$ at 95\% CL~\cite{Tristram:2021tvh}. Interestingly enough, this limitation is generically surpassed in the presence of sizable non-minimal couplings to gravity~\cite{Galante:2014ifa, Tenkanen:2017jih, Rubio:2018ogq}. 

The inclusion of non-minimal couplings to gravity explicitly breaks the well-known degeneracy between metric and Palatini formulations, making it necessary to specify the properties of the connection $\widetilde\Gamma^\rho{}_{\mu\nu}$ in order to completely define the theory under consideration. For the sake of simplicity and without lack of generality, we will assume in what follows a Palatini formulation of gravity where the connection $\widetilde\Gamma^\rho{}_{\mu\nu}$ is taken to be arbitrary but torsion-free, i.e. $\widetilde\Gamma^\rho{}_{\mu\nu}=\widetilde\Gamma^\rho{}_{\nu\mu}$. Compared to the most common metric approach, this formulation displays some interesting features. On the one hand, it does not require the introduction of the usual Gibbons–Hawking–York term to obtain the equations of motion~\cite{Ferraris1982}. On the other hand, since the metric and the connection are completely unrelated in Palatini gravity, the Ricci scalar remains invariant under Weyl transformations, simplifying the transitions among conformal frames and the analysis of the cosmological implications of the model, as we explicitly demonstrate in what follows.  

\begin{figure}[t!]
    \centering
    \includegraphics[scale=0.49]{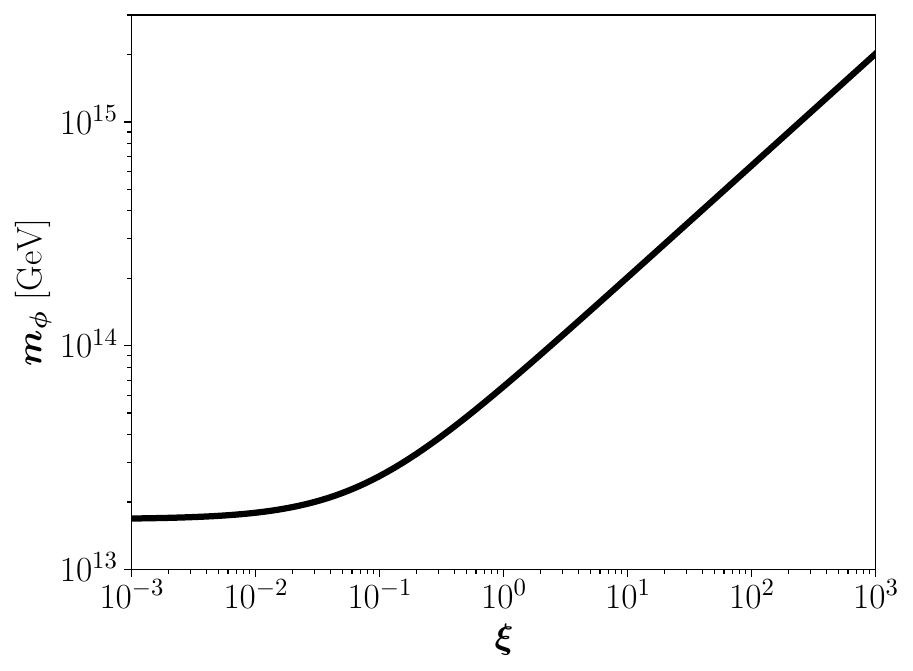}
    \caption{Dependence of the inflaton mass on the non-minimal coupling $\xi$ for $N = 50$ $e$-folds.}
    \label{fig:Ps1}
\end{figure}

\begin{figure}[t!]
    \centering
    \includegraphics[scale=0.49]{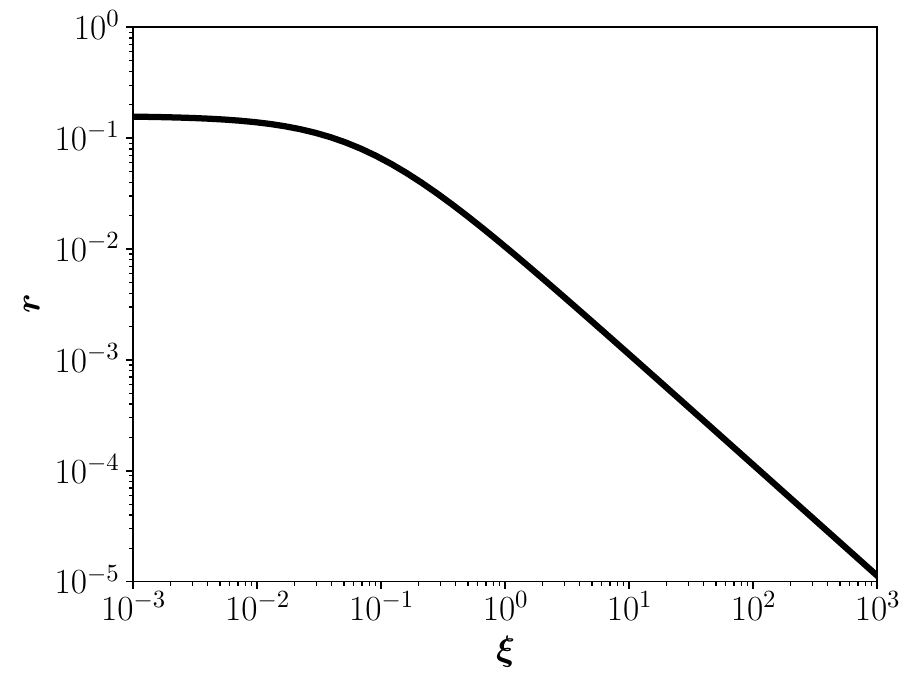}
    \includegraphics[scale=0.49]{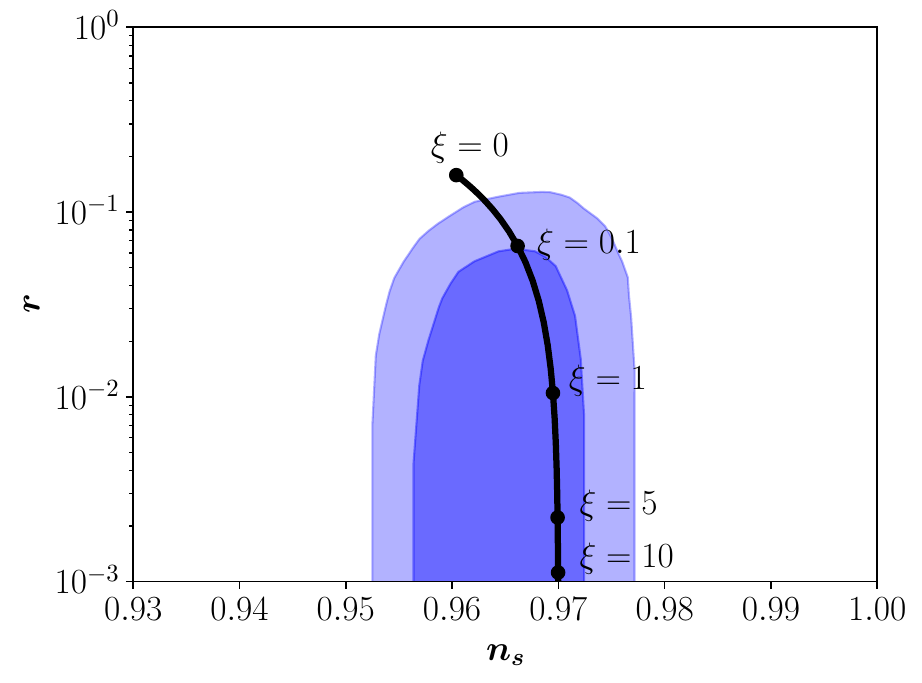}
    \caption{Dependence of the tensor-to-scalar ratio $r$ and the spectral tilt $n_s$ on the non-minimal coupling $\xi$ for $N = 50$ $e$-folds.}
    \label{fig:r-ns}
\end{figure}

The nonlinearities associated with the non-minimal coupling in Eq.~\eqref{Jframe} can be transferred to the kinetic and potential sectors of the theory by performing a Weyl transformation $\tilde g_{\mu\nu} = F^{-1}(\tilde\phi)\, g_{\mu\nu}$, which, as anticipated, affects only the metric field and its determinant. The resulting action takes the form 
\begin{equation}\label{eq:SEF}
    S_\phi = \int d^4x\, \sqrt{-g} \left[\frac{M_P^2}{2}\,R(\Gamma) +\frac{1}{2\, F(\tilde\phi)}\,
    g^{\mu\nu}\, \partial_\mu\tilde\phi\, \partial_\nu \tilde\phi -  \frac12 \frac{m_\phi^2\, \tilde \phi^2}{F^2(\tilde\phi)}
    \right],
\end{equation}
with $\Gamma$ identified now with the Levi-Civita connection, 
\begin{equation}\label{LCconnection}
    {\Gamma}^\lambda_{\alpha\beta} = \frac{1}{2}g^{\lambda\rho}\,\left(\partial_\alpha g_{\beta\rho} + \partial_\beta g_{\rho\alpha} - \partial_\rho g_{\alpha\beta}\right)\,.
\end{equation}
The non-canonical kinetic term in Eq.~\eqref{eq:SEF} can be made canonical by performing an additional field redefinition 
\begin{equation}\label{fieldred}
\frac{d\vert \phi\vert}{d\tilde\phi}= \frac{1}{\sqrt{F}}\,.
\end{equation}
Assuming the expansion of $F(\tilde\phi)$ to admit a dominant linear term\footnote{This happens in any other model of inflation that involves Planckian field excursions. This assumption requires a certain level of fine-tuning, in particular, given a general function $F(\tilde \phi)= 1+\xi \frac{\vert \tilde \phi \vert }{M_P}+\sum_n \beta_n \frac{\vert \tilde \phi \vert^n}{M_P^{n}}$ with $n\geq 2$ and sizable field excursions $\vert \tilde\phi\vert\sim {\cal O}(M_P)$, we require $\beta_n \ll \xi$.} in the field regime of interest with positive coefficient $\xi$,\footnote{The absolute value in this expression respects the assumed $\mathbb{Z}_2$ symmetry of the low energy inflaton sector, guaranteeing also a positive definite graviton propagator at all field values. On top of that, this choice ensures the absence of non-perturbative tachyonic resonance effects~\cite{Dufaux:2006ee} that would require dedicated lattice simulations, allowing us to concentrate on purely perturbative gravitational production channels in what follows.}
\begin{equation} \label{eq:Fexp}
     F(\tilde \phi) = 1 + \xi\, \frac{|\tilde\phi|}{M_P} + \mathcal{O}\left(\vert\tilde\phi\vert/M_P\right)^2\,,
\end{equation}
the integration of Eq.~\eqref{fieldred} with boundary condition $\phi(0)=0$ provides a relation  
\begin{equation}
    \vert \phi\vert = \frac{2M_P}{\xi}\left(\sqrt{1+\xi\frac{\vert\tilde \phi\vert}{M_P}}-1\right),
\end{equation}
which can be easily inverted, 
\begin{equation}
    \vert\tilde \phi\vert= \vert\phi\vert \left(1 + \frac{\xi}{4} \frac{\vert \phi\vert}{M_P}\right),
\end{equation}
to obtain
a $\phi$-dependent action 
\begin{equation} \label{eq:Sfinal}
    S_\phi = \int d^4x\, \sqrt{-g} \left[\frac{M_P^2}{2}\, R + \frac12\, g^{\mu\nu}\,\partial_\mu\phi\,\partial_\nu \phi - V(\phi)\right],
\end{equation}
with effectively \textit{non-linear $\mathbb{Z}_2$} symmetric potential
\begin{equation}\label{eq:pot-ein}
    V = \frac12\, m_\phi^2\, \vert\phi\vert^2\, \frac{ \left(1 + \frac{\xi}{4} \frac{\vert \phi\vert}{M_P}\right)^2}{ \left(1+\frac{\xi\vert \phi\vert}{2M_P}\right)^4}\,.
\end{equation}
Restricting ourselves to the asymptotic plateau-like region at large field values $\phi>0$, and dropping consequently the absolute value in all the following expressions, we obtain the potential slow-roll (SR) parameters
\begin{align}\label{eq:SR}
    \epsilon_V &\equiv  \frac{M_P^2}{2}\,\left(\frac{V'}{V}\right)^2=2\left(\frac{M_P}{\phi}\right)^2\left(1+\frac{1}{2}\frac{\xi\phi}{M_P}\right)^{-2}\left(1+\frac{1}{4} \frac{\xi\phi}{M_P}\right)^{-2},\\
    \eta_V &\equiv  M_P^2\,\frac{V'}{V}= \left[1-\frac{3}{2} \frac{\xi  \phi }{M_P}-\frac{3}{8}\left(\frac{\xi\phi}{M_P}\right)^2\right]\epsilon_V\,,\label{eq:SR1}
\end{align}
and the number of $e$-folds of inflation
\begin{equation} \label{eq:Ne}  
    N = \frac{1}{M_P^2} \int^{\phi_*}_{\phi_{\rm end}} \frac{V}{V'}\, d\phi = \frac14 \left(\frac{\phi}{ M_P}\right)^2\left(1+\frac{\xi}{4} \frac{\phi}{M_P}\right)^2\Bigg\vert^{\phi_*}_{ \phi_{\rm end}} \,,
\end{equation}
between the field value $\phi_*$ at which the pivot scale $k_*$ exited the horizon during inflation and the corresponding one at the very end of inflation, 
\begin{equation}\label{eq:phi-end}
    \phi_{\rm end} \simeq \frac{2^{3/2}\,M_P}{(\sqrt{2}\, \xi)^{2/3}}\left(1- \frac{ (\sqrt{2}\,\xi)^{1/3}-1/3}{(\sqrt{2}\,\xi)^{2/3}} \right)\,,
\end{equation}
where SR is completely violated, that is $\epsilon_V(\phi_{\rm end})\simeq 1$. Neglecting the small contribution of the lower limit in Eq.~\eqref{eq:Ne} and solving for $\phi_*$, 
\begin{equation}
    \phi_*(N)= \frac{2\,M_P}{\xi}\left(\sqrt{1+ 2\, \xi\,N^{1/2}}-1\right)\,,
\end{equation}
we can now express the SR parameters \eqref{eq:SR} and \eqref{eq:SR1} as a function of the number of $e$-folds of inflation,
\begin{equation}
    \epsilon_V \simeq \frac{1}{4\,\xi \,N^{3/2}}  +{\cal O}\left(\frac{1}{\xi^2  N^{2}}\right), \hspace{10mm}
    \eta_V \simeq -\frac{3}{4\,N} + \frac{5}{8\,\xi\,N^{3/2}} +{\cal O}\left(\frac{1}{\xi^2  N^{2}}\right)\,.
\end{equation}
This allows us to determine the amplitude of the amplitude of the primordial spectrum of curvature perturbations, the associated spectral tilt $n_s = 1 + 2\, \eta_V - 6\, \epsilon_V$ and the tensor-to-scalar ratio $r=16\,\epsilon_V$, 
\begin{equation}
    \mathcal{P} \simeq  \frac{m_\phi^2}{12\,\pi^2 M_P^2} \left(\frac{N^{3/2}}{\xi}\right),
    \hspace{10mm}
    n_s = 1-\frac{3}{2 N} - \frac{1}{4\,\xi\,N^{3/2}}\,, \hspace{15mm}
    r = \frac{4}{\xi\,N^{3/2}} \,.  \label{eq:nsr}
\end{equation}
The observed spectral amplitude $\mathcal{P} \simeq 2.1 \times 10^{-9}$~\cite{Planck:2018jri} determines the inflaton mass,
\begin{equation}\label{eq:mphi-N}
    m_\phi^2\simeq 12\,\pi^2\, \mathcal{P}\, M_P^2\,\left(\frac{\xi}{N^{3/2}}\right)\,,
\end{equation}
which, as shown in Fig.~\ref{fig:Ps1}, turns out to exceed generically the unification scale $m_\phi\sim {\cal O}(10^{13})$~GeV associated with quadratic chaotic models of inflation, with larger values of the non-minimal coupling $\xi$ leading to larger inflaton masses. Requiring $m_\phi$ to stay sub-Planckian results in an upper bound for $\xi$, namely 
\begin{equation}
    \xi \lesssim 1.5 \times 10^9 \left(\frac{N}{50}\right)^{3/2} \,. 
\end{equation}
%%%%%%%%%%%%%%%%%%%%%%%%%%%%%%%%%%%%%%%%%%%%%%%%%%%
%%%%%%%%%%%%%%%%%%%%%%%%%%%%%%%%%%%%%%%%%% 
The compatibility of the spectral tilt and the tensor-to-scalar ratio with the CMB observations due to Planck~\cite{Planck:2018jri} is displayed in Fig.~\ref{fig:r-ns}. Nonvanishing values of $\xi$ result generically in small tensor-to-scalar ratios, which, as anticipated, are fully compatible with Planck data for a fixed number of $e$-folds $N=50$, a fiducial value to be assumed in what follows. In some cases, the predicted tensor-to-scalar ratios are also well within the reach of current or future planned experiments such as BICEP3~\cite{Wu:2015oig}, LiteBIRD~\cite{Matsumura:2013aja} and the Simons Observatory~\cite{SimonsObservatory:2018koc}.
%%%%%%%%%%%%%%%%%%%%%%%%%%%%%%%%%%%%%%%%%%%%%%%%%%%
%%%%%%%%%%%%%%%%%%%%%%%%%%%%%%%%%%%%%%%%%% 

%%%%%%%%%%%%%%%%%%%%%%%%%%%%%%%%%%%%%%%%%
\section{Heating with a Linear Non-minimal Coupling}
\label{sec:heating}
%%%%%%%%%%%%%%%%%%%%%%%%%%%%%%%%%%%%%%%%%
In this section we explore the aftermaths of the inflaton decay induced by the linear non-minimal coupling, neglecting for simplicity additional decay channels potentially allowed by our symmetry assumptions (see the discussion at the end of Appendix~\ref{sec:decay-branching} for further clarification). We begin with the production of the SM bath from the decay of the inflaton field, which is necessary for heating. We then discuss the production of DM and the dynamical generation of the BAU, for which we introduce new physics states. Inevitably, in all cases the yield is proportional to the square of the non-minimal coupling. We emphasize that all our computations consider perturbative 2-body decay of the inflaton condensate, neglecting in particular non-perturbative production effects. This approximation is justified by the quadratic character on the inflationary potential around its minimum and the Planck-suppressed character of the interactions induced by the non-minimal coupling to gravity. The former aspects constitute, in fact, a remarkable difference from scenarios involving higher monomial potentials, {\it viz.,} $V(\phi)\propto\phi^k$, where, despite what is usually assumed in the literature~\cite{Haque:2022kez, Co:2022bgh, Barman:2022qgt, Clery:2023mjo}, non-perturbative effects cannot be generically ignored, leading almost universally to a radiation-like equation-of-state parameter~\cite{Lozanov:2016hid, Lozanov:2017hjm, Bettoni:2021zhq, Laverda:2023uqv, Garcia:2023dyf, Barman:2023ktz} in clear contrast to the value $w = (k - 2)/(k + 2)$ obtained in the homogeneous approximation.

%%%%%%%%%%%%%%%%%%%%%%%%%%%%%%%%%%%%%%%%%
\subsection{Producing the standard model bath}
\label{sec:radiation}
%%%%%%%%%%%%%%%%%%%%%%%%
At the end of inflation, the potential energy of $\phi$ becomes comparable to its kinetic energy counterpart, leading effectively to a Hubble parameter of order $H_I \equiv H(\phi_{\rm end}) \simeq V^{1/2}(\phi_\text{end})/M_P \sim 10^{13}$~GeV, with a very mild dependence on the non-minimal coupling $\xi$. In order to recover the usual hot Big Bang evolution, this large energy contribution must be transferred to the SM degrees of freedom, heating the Universe and ensuring the onset of radiation domination. For the sake of generality, we will assume that the inflaton field can also decay into new-physics (NP) states beyond the SM, with a suppressed branching fraction $\br$. The evolution of the SM energy density $(\rR)$ and the inflaton $(\np)$ and NP ($\nnp$) number densities can be tracked with the set of coupled Boltzmann differential equations~\cite{Drees:2017iod, Arias:2019uol}
\begin{align}
    &\frac{d\np}{dt} + 3\, H\, \np = -\Gp\, \np\,,  \label{eq:beq1}\\
    &\frac{d\rR}{dt} + 4\,H\, \rR = \left(1-\br\right) \Gp\, \np\, \mphi\,,  \label{eq:beq2}\\
    &\frac{d\nnp}{dt} + 3\,H\, \nnp = 2\, \br\, \Gp\, \np\,,  \label{eq:beq3}
\end{align}
with $\Gp$ and $\rp \equiv \np\, \mphi$ the total decay width and energy density of the nonrelativistic inflaton, respectively, and 
\begin{equation}
    H \simeq \sqrt{\frac{\rR + \np\, \mphi}{3\, M_P^2}}
\end{equation}
the Hubble expansion rate. In the following, the new state will be identified with the DM or the RHN responsible for leptogenesis.
The SM radiation energy density as a function of the SM bath temperature $T$ is given by
\begin{equation}
    \rR(T) = \frac{\pi^2}{30}\, \gs(T)\, T^4\,,
\end{equation}
where $\gs(T)$ corresponds to the SM relativistic degrees of freedom contributing to $\rR$~\cite{Drees:2015exa}. For the sake of simplicity, we restrict ourselves to instantaneous thermalization within the SM sector. More precisely, we will assume that the interaction rate between SM particles significantly exceeds the inflaton decay rate~\cite{Kolb:2003ke}.

The heating temperature $\Trh$ can be defined as the temperature of the SM bath at which the equality $\Gp = H(\Trh)$ occurs and corresponds to
\begin{equation} \label{eq:trh}
    \Trh^2 = \frac{3}{\pi} \sqrt{\frac{10}{\gs}}\, M_P\, (1-\br)\, \Gp\,.
\end{equation}
As the inflaton decay is not instantaneous, the maximum temperature~\cite{Chung:1998rq, Giudice:2000ex, Barman:2021ugy}
\begin{equation} \label{eq:Tmax}
    \Tmax^4 = \frac{60}{\pi^2\,\gs} \left(\frac38\right)^{8/5} \left(1-\br\right)\, M_P^2\, \Gp\, H_I
\end{equation}
reached by the SM bath can be much higher than $\Trh$. 

Assuming a perturbative decay of the inflaton field, the partial decay widths into final-state particles of mass $m$, different spins, and a {\it single} degree of freedom is
\begin{equation} \label{eq:Gamma}
    \Gamma_{\phi\to ii}= 
    \begin{dcases}
         \frac{\xi^2}{32\,\pi}\,\frac{\mphi^3}{M_P^2}\,\sqrt{1-4\,y^2}   &\text{for scalars,}\\
        \frac{\xi^2}{32\,\pi}\,\frac{\mphi^3}{M_P^2}\,y^2\,\left(1-4\,y^2\right)^{3/2}   &\text{for fermions,}\\
        \frac{\xi^2}{128\,\pi}\,\frac{\mphi^3}{M_P^2}\,\sqrt{1-4\,y^2}\,\left(1-4\,y^2+12\,y^4\right)   &\text{for vectors}\,,
    \end{dcases}
\end{equation}
with $y \equiv m/\mphi$. Details of the calculation are reported in Appendix~\ref{sec:decay-branching}.
At high temperatures before the electroweak symmetry breaking, the SM particles are massless and the SM contains 4 scalar, 24 vector and 90 fermionic degrees of freedom. Therefore, the total decay width into SM final states becomes
\begin{equation} \label{eq:Gammatot}
    \Gp \simeq \frac{5}{16 \pi}\, \xi^2\, \frac{\mphi^3}{M_P^2}\,.
\end{equation}
Using this perturbative expression for $\Gp$ and the inflaton mass in Eq.~\eqref{eq:mphi-N}, we get 
\begin{equation}
    \Trh \simeq 3.1\times 10^{12}~\text{GeV}\,\left(\frac{\xi}{10}\right)^{7/4},
\end{equation}
and
\begin{equation}
    \Tmax \simeq 5.4\times 10^{13}~\text{GeV} \left(\frac{\xi}{10}\right)^{5/6},
\end{equation}
for $\br =0$.
%%%%%%%%%%%%%%%%%%%%%%%%%%%%
\begin{figure}[t!]
    \centering
    \includegraphics[scale=0.37]{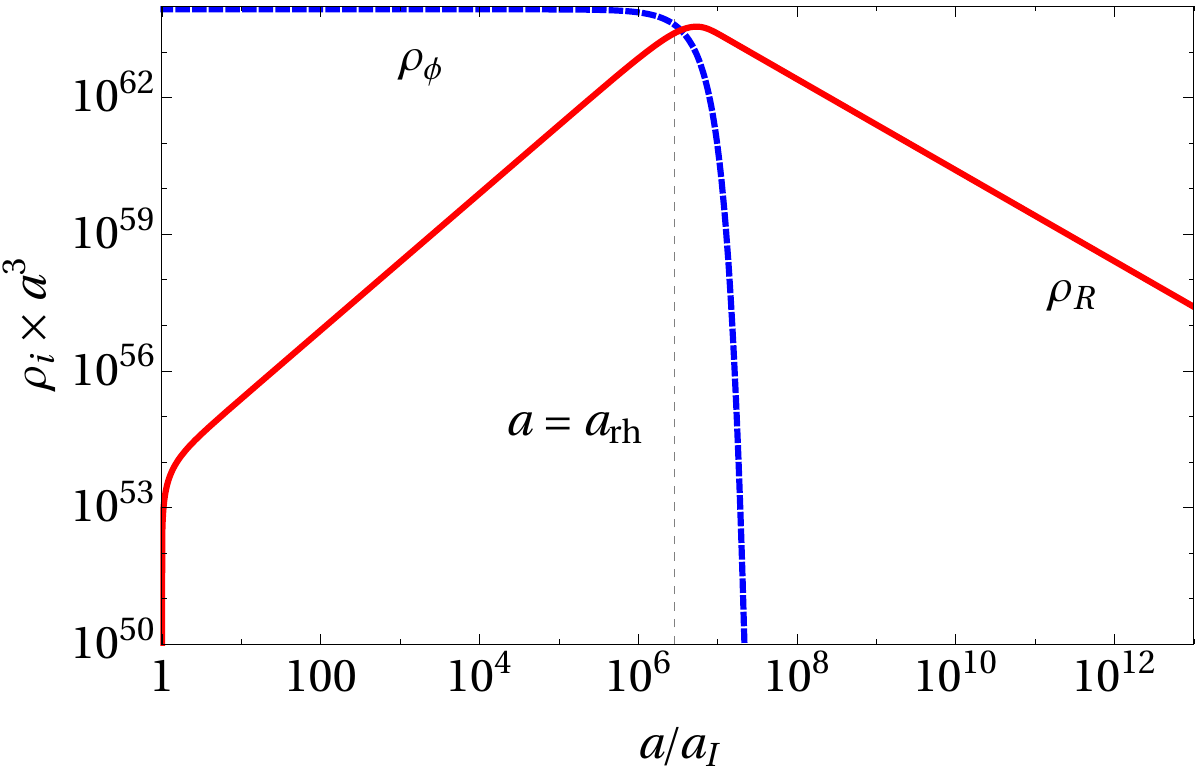}
    \includegraphics[scale=0.37]{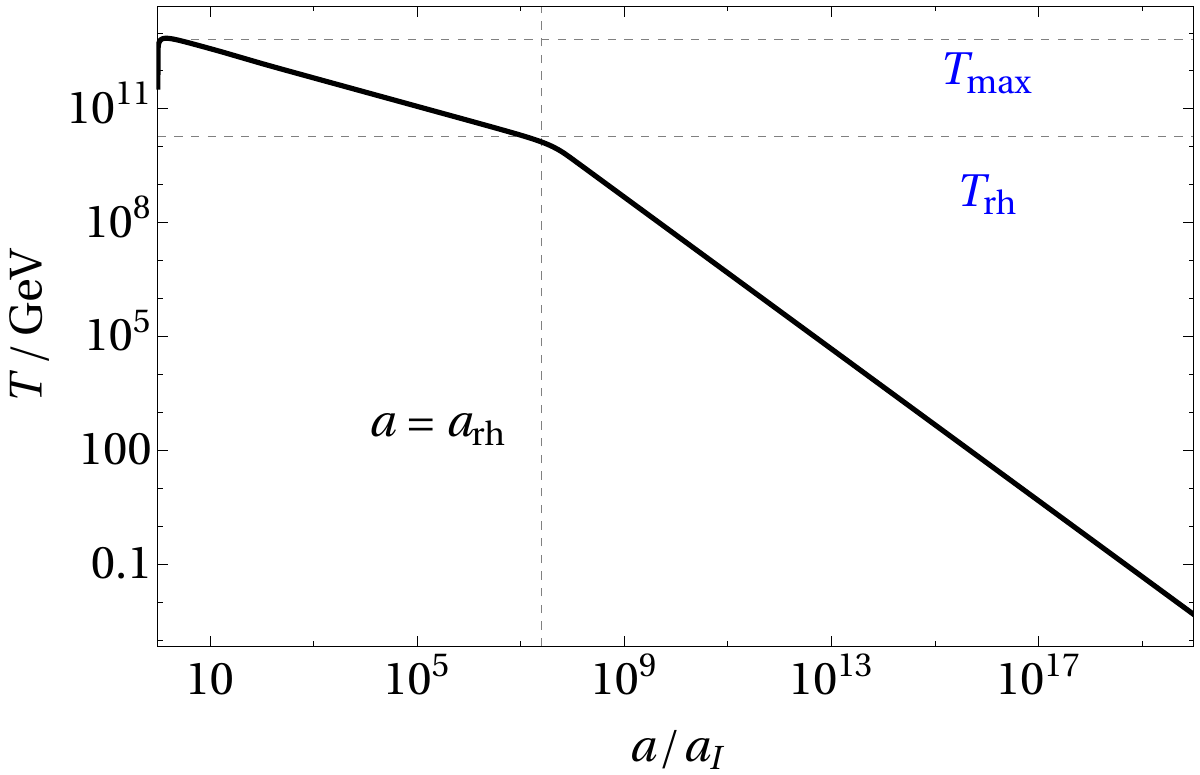}
    \caption{Left: Evolution of the radiation (red) and inflaton (blue) energy densities with the scale factor. Right: Bath temperature as a function of the scale factor. In both panels, we fix $\br = 0$ and $\Gp \simeq 5 \times 10^3$~GeV, which, in the perturbative regime, corresponds to $\xi \simeq 1$.
    }
    \label{fig:energy}
\end{figure} 
%%%%%%%%%%%%%%%%%%%%%%%%%%%%%%%%%%%%%%%%%%
The evolution of the inflaton (red) and the SM radiation (blue) energy densities as a function of the cosmic scale factor $a$ is displayed in the left panel of Fig.~\ref{fig:energy}, for $\Gp \simeq 5$~TeV ($\xi\simeq 1$) and only considering decays into the SM, i.e., $\br = 0$. The vertical dashed line corresponds to $a = \arh \equiv a(\Trh)$. During heating, that is, in the range $a_I < a < \arh$ (with $a_I$ corresponding to the scale factor and the end of inflation / beginning of heating), $\rp(a) \propto a^{-3}$ while $\rR(a) \propto a^{-3/2}$, as it is not just a free radiation component but rather a one sourced from the decay of the inflaton field.

In addition, in the right panel, the evolution of the SM temperature as a function of the scale factor is shown. Horizontal dashed lines correspond to $T = \Tmax$ and $T = \Trh$ and delimit the heating duration. At the beginning of heating, the bath temperature rapidly increases as a result of the non-instantaneous decay of the inflaton field, reaching a temperature $\Tmax \sim 8\times 10^{12}$~GeV. During heating, the temperature decreases as $T(a) \propto a^{-3/8}$, until $\Trh \sim 6\times 10^{10}$~GeV. Once SM radiation dominates the energy density of the Universe, it becomes a free radiation fluid and its energy density drops to $\rR(a) \propto a^{-4}$, corresponding to $T(a) \propto a^{-1}$.

%%%%%%%%%%%%%%%%%%%%%%%%%%%%%%%%%%%%%%%%%%
\begin{figure}[t!]
    \centering
    \includegraphics[scale=0.62]{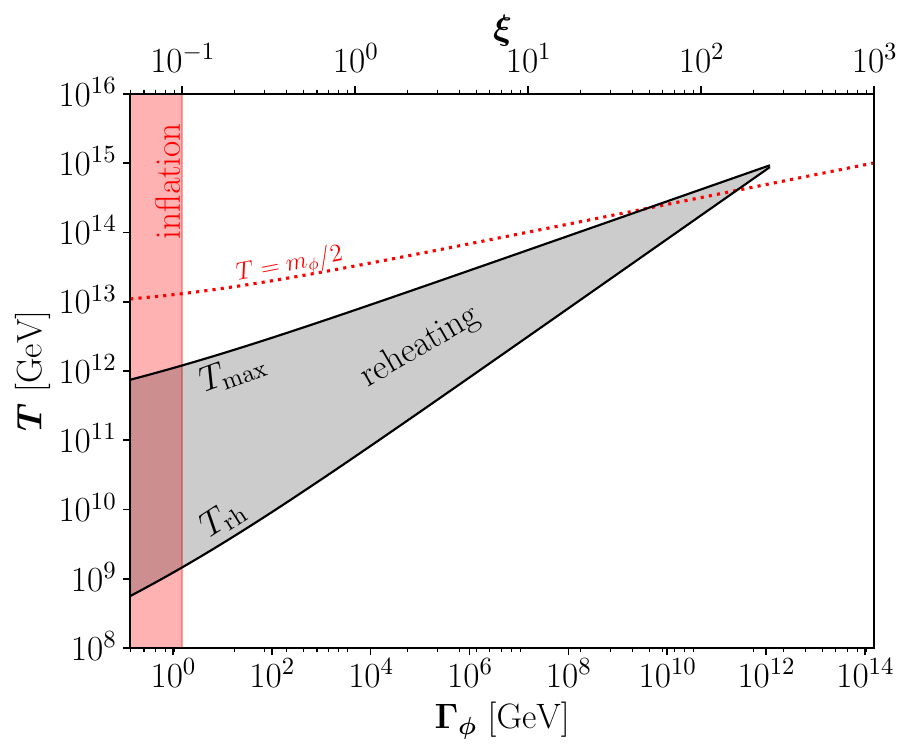}
    \caption{Values of the heating temperature $\Trh$ and the maximum temperature $\Tmax$ as a function of the non-minimal coupling $\xi$ for $\br = 0$. The gray-shaded region $(\Trh \leq T \leq \Tmax)$ corresponds to the heating epoch. 
    }
    \label{fig:bound}
\end{figure} 
%%%%%%%%%%%%%%%%%%%%%%%%%%%%%%%%%%%%%%%%%%
Figure~\ref{fig:bound} shows the values for $\Trh$ and $\Tmax$ as a function of $\Gp$ (or $\xi$ in the perturbative regime), assuming again $\br = 0$. The region between $\Tmax > T > \Trh$ corresponds to the heating era. The minimal value of the non-minimal coupling $\xi \gtrsim 0.1$ comes, mainly, from the inflationary tensor-to-scalar ratio; see Figs.~\ref{fig:r-ns} and \eqref{eq:nsr}. Additionally, an upper bound 
\begin{equation}
    \xi \lesssim 250
\end{equation}
on the non-minimal coupling appears by demanding $\Tmax > \Trh$ or equivalently 
\begin{equation}\label{eq:xibound}
    \Gp < \frac23 \left(\frac38\right)^{8/5} \frac{H_I}{1 - \br}\,.
\end{equation}
This corresponds to a minimum bound $r \gtrsim 10^{-4}$ on the tensor-to-scalar ratio, that is, within the reach of future and planned CMB experiments~\cite{Wu:2015oig, Matsumura:2013aja, SimonsObservatory:2018koc}. Above the red dotted line, corresponding to $T = \mphi/2$, the SM bath had an energy high enough to generate inflatons through inverse decays. However, because of the high inflaton number density during heating, this process is subdominant.\footnote{In this case, additionally to the decay term $\Gp\, \np$ in Eqs.~\eqref{eq:beq1} to~\eqref{eq:beq3}, the production out of the SM bath has to be included and therefore $\Gp\, \np \to \Gp\, (\np - \np^\text{eq})$, where $\np^\text{eq}(T)$ corresponds to the equilibrium number density of the inflaton.}

%%%%%%%%%%%%%%%%%%%%%%%%%%%%%%%%%%%%%%%%%
\subsection{Dark matter from inflaton decay}
\label{sec:DM}
%%%%%%%%%%%%%%%%%%%%%%%%%%%%%%%%%%%%%%%%%
In this section, we discuss the prospect of DM production from the decay of the non-minimally coupled inflaton field. We consider the simple scenario in which the inflaton condensate decays into a pair of DM particles of arbitrary spin. The evolution of the DM number density $\ndm$ can be tracked by solving Eqs.~\eqref{eq:beq1} to~\eqref{eq:beq3}, where $\nnp$ becomes $\ndm$.
Equation~\eqref{eq:beq3} can be conveniently rewritten in terms of the scale factor $a$ and the comoving number density $N \equiv \ndm\, \times a^3$ as
\begin{equation} \label{eq:dm-num}
    \frac{dN}{da} = 2\, \br\, \frac{a^2\, \Gp}{H(a)}\, n_\phi(a) \simeq 6\, \br\, \frac{M_P^2\, \Gp^2}{\mphi}\, \arh^{3/2}\, a^{1/2}\,,
\end{equation}
where the scaling of the inflaton number density $n_\phi(a) \propto a^{-3}$ during heating was used, with $\arh \equiv a(\Trh)$. It is interesting to note that, due to the nature of gravitational couplings, the branching fraction $\br$ is not a free parameter and depends only on the spin (and mildly the mass) of the decaying particle. The branching of the inflaton field into a couple of DM particles (with a single degree of freedom) in the final state follows from Eqs.~\eqref{eq:Gamma} and~\eqref{eq:Gammatot}, and is given by
\begin{equation}
    \br \simeq 
    \begin{dcases}
        \frac{1}{11}-\frac{20}{121} \, y^2 + \mathcal{O}[y^4] &\text{for scalars,}\\
        \frac{1}{10} \, y^2 + \mathcal{O}[y^4] &\text{for fermions,}\\
       \frac{1}{41}-\frac{240}{1681} \, y^2 + \mathcal{O}[y^4] &\text{for vectors},
    \end{dcases}
\end{equation}
for $m\ll m_\phi$. We emphasize that due to the democratic gravitational interaction strength, the branching ratio is independent of the non-minimal coupling.

Next, we note that Eq.~\eqref{eq:dm-num} admits an analytical solution 
\begin{equation} \label{eq:Nrh}
    N(\arh) \simeq 4\, \br\, \frac{M_P^2\, \Gp^2}{\mphi}\, \arh^3 \left[1-\left(\frac{\Trh}{\Tmax}\right)^4\right]\,,
\end{equation}
where we have assumed that there is no initial population of DM at the end of inflation, that is, $\ndm(a_I) \simeq  0$.
In addition, one can define the DM yield $Y \equiv \ndm/s$, where
\begin{equation}
    s(T) = \frac{2 \pi^2}{45}\, \gss(T)\, T^3
\end{equation}
is the SM entropy density, and $\gss(T)$ is the number of relativistic degrees of freedom contributing to the SM entropy~\cite{Drees:2015exa}.
The value of the DM yield at present corresponds to the value at the end of the heating and is given by 
\begin{equation}
    Y(\arh) \simeq \frac{N(\arh)}{\arh^3\, s(\Trh)} \simeq 4\, \br\, \frac{M_P^2\, \Gp^2}{\mphi\, s(\Trh)} \left[1-\left(\frac{\Trh}{\Tmax}\right)^4\right],
\end{equation}
which in the perturbative case corresponds to
\begin{equation}\label{eq:yld}
    Y(\arh) \simeq \frac{\xi}{\gss(\Trh)^{1/4}}\,\sqrt{\frac{m_\phi}{M_P}} \times
    \begin{dcases}
        1 &\text{ for scalar DM,}\\
        \left(\frac{\mdm}{\mphi}\right)^2 &\text{ for fermionic DM,}\\
        \frac12 \sqrt{\frac{11}{41}} &\text{ for vector DM,}
    \end{dcases}
\end{equation}
featuring a linear dependence on the nonminimal coupling $\xi$.
To match the entire observed abundance, the DM yield must be fixed so that $\mdm\, Y(\arh) = \Omega_\text{DM} h^2 \frac{1}{s_0}\, \frac{\rho_c}{h^2} \simeq 4.3 \times 10^{-10}$~GeV, where $\mdm$ is the DM mass, $\rho_c \simeq 1.1 \times 10^{-5}~h^2$~GeV/cm$^3$ is the critical energy density, $s_0 \simeq 2.9 \times 10^3$~cm$^{-3}$ is the entropy density at present, and $\Omega_\text{DM} h^2 \simeq 0.12$~\cite{Planck:2018jri}.

\begin{figure}[t!]
    \centering
    \includegraphics[scale=0.65]{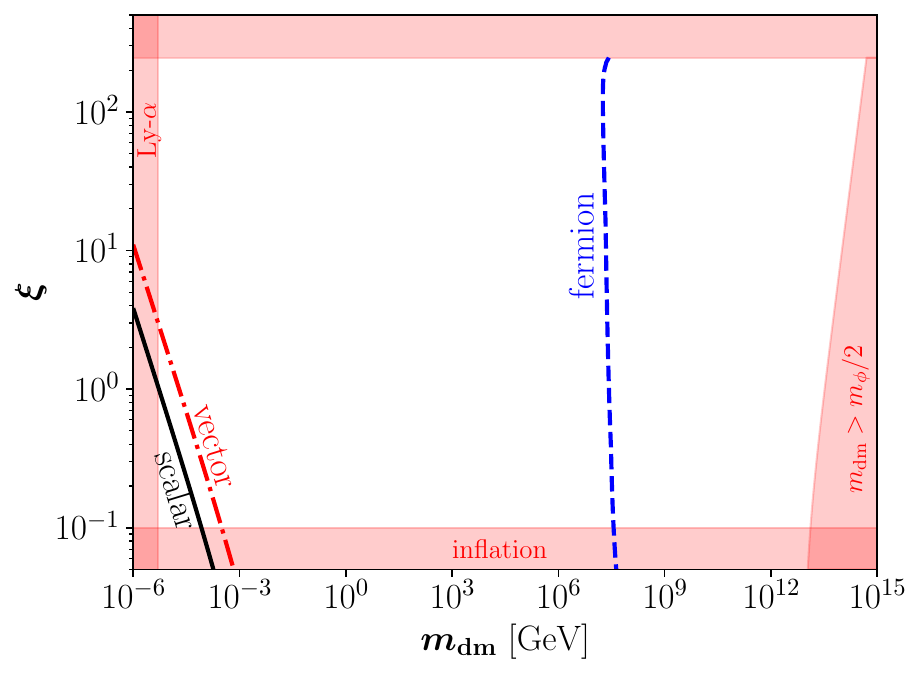}
    \caption{Magnitude of the non-minimal coupling $\xi$ needed to fit the entire observed DM abundance, for different DM spins. The red-shaded regions are disallowed by BBN (bottom), Lyman-$\alpha$ (left), kinematical condition $\mdm > \mphi/2$ (right) and nonviable heating (top).}
    \label{fig:relic}
\end{figure}

In Fig.~\ref{fig:relic}, the values of $\xi$ required to make up the entire DM relic abundance at present are shown, as a function of the mass of the DM, for different spins of the DM. Using Eq.~\eqref{eq:yld}, together with Eq.~\eqref{eq:mphi-N}, one finds that for bosonic cases the yield $Y(\arh) \propto \xi^{5/4}$, while for the fermionic case, because of helicity suppression, $Y(\arh) \propto \mdm^2\, \xi^{1/4}$. For the same reason, for fermionic DM, $\mdm$ varies very steeply with $\xi$, as compared to bosonic cases. As a result, for a given $\xi$, the fermionic DM needs to be heavier than the bosonic one in order to produce the right amount of relic.
The DM mass required to fit the whole observed abundance is therefore
\begin{equation}
    \mdm \simeq
    \begin{dcases}
            5.3\times 10^{-6}~\text{GeV} \times \xi^{-5/4} &\text{ for scalar DM,}\\
            2.7\times 10^7~\text{GeV} \times \xi^{-1/12} &\text{ for fermionic DM,}\\
            1.9\times 10^{-5}~\text{GeV} \times \xi^{-5/4} &\text{ for vector DM},
    \end{dcases}
\end{equation}
for $\gss(\Trh) = 106.75$. 
In Fig.~\ref{fig:relic}, we also show constraints on the viable parameter space from Lyman-$\alpha$ flux power spectra on a warm DM mass~\cite{Narayanan:2000tp, Viel:2005qj, Baur:2015jsy, Irsic:2017ixq, Palanque-Delabrouille:2019iyz, Garzilli:2019qki} that allows $\mdm \gtrsim 4$~keV~\cite{Ballesteros:2020adh, DEramo:2020gpr}, on-shell decay of the inflaton that requires $\mdm < \mphi/2$, and successful heating followed by inflation; cf. Fig.~\ref{fig:bound}. Once these constraints are taken into account, the allowed mass range for scalar and vector DM turns out to be $4~\text{keV} \lesssim \mdm \lesssim 1$~MeV, while for fermionic DM $\mdm \simeq 10^7$~GeV.
%%%%%%%%%%%%%%%%%%%%%%%%%%%%%%%%%%%%%%%%%%%%%%%%%%%
%%%%%%%%%%%%%%%%%%%%%%%%%%%%%%%%%%%%%%%%%% 

Before closing, we would like to mention that, apart from direct decay of the inflaton field into DM final states, pure gravitational production of DM unavoidably takes place from the 2-to-2 scattering of the bath particles via $s$-channel mediation of massless graviton~\cite{Garny:2015sjg, Tang:2017hvq, Garny:2017kha, Bernal:2018qlk, Barman:2021ugy, Barman:2021qds}. However, since the density rate of DM production in such a scenario scales as $\gamma(T) \propto T^8/M_P^4$, the gravitational UV freeze-in is subdominant.

%%%%%%%%%%%%%%%%%%%%%%%%%%%%%
\subsection{Leptogenesis from inflaton decay}
\label{sec:lepto}
%%%%%%%%%%%%%%%%%%%%%%%%%%%%%%%
Aside DM, the dynamical generation of the observed BAU demands the introduction of new physics. To be more accurate, neutrino masses and mixings require at least two (heavy) right-handed neutrino (RHN) states to realize the seesaw mechanism~\cite{Minkowski:1977sc, GellMann:1980vs, Yanagida:1979as, Mohapatra:1979ia}. One of these, if produced and remains out-of-equilibrium until its decay, can leave a nonzero lepton asymmetry. This asymmetry in the leptonic sector can eventually be converted into an asymmetry in the baryonic sector following the well-known mechanism of leptogenesis~\cite{Fukugita:1986hr, Giudice:2003jh, Davidson:2008bu}. In the present context, such a framework can be realized by considering inflaton decays into a pair of RHNs, which further undergoes CP-violating out-of-equilibrium decays into a SM lepton and a Higgs.

As a concrete example, we introduce SM gauge singlet RHNs $N_i$ (with $i= 1$, 2, 3) with an interaction Lagrangian density of the form 
\begin{equation} \label{eq:RHN-lgrng}
    \mathcal{L} \supset -\frac12\, m_N\, \overline{N^c}\, N - y_N\, \overline{N}\, \widetilde{H}^\dagger\, L + {\rm H.c.}\,,
\end{equation}
ignoring the generational indices, where $L$ is the SM lepton doublet, $H$ the SM complex Higgs doublet $(\widetilde H \equiv i\, \sigma^2\, H^\star$, $\sigma^2$ is the Pauli spin matrix), and $y_N$ a Yukawa coupling. Additionally, $m_{N_i}$ are the Majorana masses, assumed to be hierarchical $m_{N_1} \ll m_{N_{2,3}}$. Note that the trilinear Yukawa term in Eq.~\eqref{eq:RHN-lgrng} is responsible for generating light neutrino masses via the Type-I seesaw mechanism.

Due to the democratic coupling of the inflaton field to all SM and NP fields, it inevitably decays into a pair of such RHNs, which subsequently undergo a CP-violating decay into a SM lepton and a Higgs doublet via their Yukawa interaction, producing a non-zero lepton asymmetry. The produced lepton asymmetry is eventually converted to baryon asymmetry via electroweak sphalerons. The final BAU can then be estimated via~\cite{Harvey:1990qw, Davidson:2008bu} 
\begin{equation} \label{eq:yb}
    Y_B^0 = \frac{28}{79}\, |\epsilon_{\Delta L}|\,  Y_{N_1}(\arh)\,,
\end{equation}
where~\cite{Kaneta:2019yjn, Co:2022bgh}
\begin{equation} \label{eq:cp}
   | \epsilon_{\Delta L} |\simeq \frac{3\, \delta_{\rm eff}}{16\,\pi}\, \frac{m_{N_1}\, m_{\nu,\text{max}}}{v^2}\,,  
\end{equation}
is the lepton asymmetry (see Appendix~\ref{sec:cp} for details), $\langle H\rangle \equiv v \simeq 174$~GeV is the SM Higgs vacuum expectation value, $\delta_{\text{eff}}$ is the effective CP violating phase in the neutrino mass matrix with $0 \leq \delta_{\rm eff} \leq 1$, and we take $m_{\nu,\text{max}} \simeq 0.05$~eV as the heaviest active neutrino mass~\cite{Harvey:1990qw}. The RHN yield at the end of heating is given by the fermionic part of Eq.~\eqref{eq:yld}.
For the decay of heavier RHNs $N_{2,3}$, we consider lepton-number-violating interactions of $N_1$ rapid enough to wash out the lepton-number asymmetry originated by the other two. Therefore, only the CP-violating asymmetry from the decay of $N_1$ survives and is relevant for leptogenesis.  

%%%%%%%%%%%%%%%%%%%%%%%%%%%%%%%%%%%%%%%%%%%%%%%%%%%
\begin{figure}[t!]
    \centering
    \includegraphics[scale=0.65]{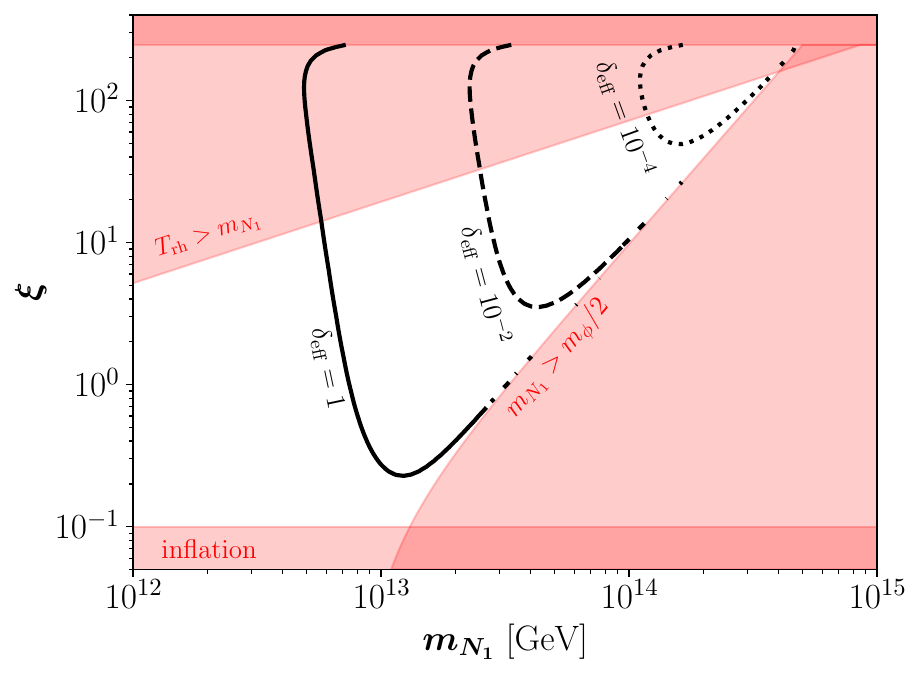}
    \caption{Contours corresponding to the observed BAU for different choices of CP-violating phases $\delta_{\rm eff}$, as shown with different patterns. The shaded regions are prohibited from inflation [cf. Fig.~\ref{fig:r-ns}], maximum temperature during heating [cf. Fig.~\ref{fig:bound}], kinematical constraint on 2-body decay and non-thermal leptogenesis that requires $m_{N_1}>\Trh$.}
    \label{fig:yB}
\end{figure} 
%%%%%%%%%%%%%%%%%%%%%%%%%%%%%%%%%%%%%%%%%%
To match the observed BAU at present, it is required to have $Y_B^0 \simeq 8.7 \times 10^{-11}$~\cite{Planck:2018vyg}.
Away from the kinematical thresholds, this implies that $m_{N_1} \propto \xi^{-1/8}$. In Fig.~\ref{fig:yB} we show with black lines the effective CP violating phase $\delta_{\text{eff}}$ required to fit the data, in the $[m_{N_1},\, \xi]$ plane. The contour shows a cutoff point at $m_{N_1}\simeq \mphi/2$, which is the kinematical threshold for 2-body decay (red area on the right). To avoid the washout of the produced asymmetry, one also needs to ensure that the production is nonthermal,\footnote{To realize non-thermal leptogenesis {\it during} heating, we compute the thermalization rate $\Gamma_{\rm th} \simeq y_N^2\,T/ (8\,\pi)$, with $y_N \simeq m_{\nu,\text{max}}\,m_{N_1} / v^2$, and compare it with the Hubble rate. We find that $\Gamma_{\rm th}$ stays below the Hubble rate $H(T) \simeq \left(T/\Tmax\right)^4\, H_I$ during heating for $\mathcal{O}(10^{-1}) \lesssim \xi \lesssim \mathcal{O}(10^2)$.} which requires $m_{N_1} > \Trh$ (shaded red on top). Therefore, within the white area corresponding to $5 \times 10^{12}~\text{GeV} \lesssim m_{N_1} \lesssim 5 \times 10^{14}$~GeV and $ 10^{-1} \lesssim \xi \lesssim 2 \times 10^2$, the BAU can be reproduced successfully.

%%%%%%%%%%%%%%%%%%%%%%%%%%%%%%%%%%%%%%%%%%%%%%%%%%%%%%%%%%
\section{Conclusions} \label{sec:concl}
%%%%%%%%%%%%%%%%%%%%%%%%%%%%%%%%%%%%%%%%%%%%%%%%%%%%%%%%%%
The precise mechanism of heating after inflation remains largely unknown, opening up different possibilities of production of the Standard Model (SM) content, along with new physics species once cosmic inflation ends. In this paper we discuss one such possibility by considering a {\it linear} coupling between the inflaton field and gravity. 
Such a non-minimal coupling triggers the decay of the inflaton condensate into pairs of all particles in the SM and beyond. Contrary to the widely discussed gravitational heating scenario, mediated by graviton exchange, in the present case one can have successful inflation together with heating for a {\it quadratic} inflaton potential, i.e., the simplest chaotic inflation scenario. We find, in order to adhere to the Planck data and reheat the Universe prior to the onset of BBN, the non-minimal coupling needs to be $\mathcal{O}(10^{-1}) \lesssim \xi \lesssim \mathcal{O}(10^2)$. 

We extend our discussion to the production of new-physics states. On the one hand, the whole observed dark-matter (DM) abundance can be successfully fitted for different spins. In particular, fermionic DM must have a mass $\mdm \sim 10^7$~GeV, while bosonic DM (scalar or vector) must be in the keV to MeV range.
On the other hand, we also discuss the generation of baryon asymmetry of the Universe via nonthermal leptogenesis, due to the CP-violating decay of a heavy right-handed neutrino produced from the inflaton decay. The observed baryon asymmetry, along with the light neutrino masses via the type-I seesaw mechanism, can be produced from out-of-equilibrium decay of a heavy right-handed neutrino in the mass range $10^{12}~\text{GeV} \lesssim m_{N_1} \lesssim 10^{15}$~GeV.

All in all, we have demonstrated that, contrary to the case of a quadratic non-minimal coupling of the inflaton to gravity discussed in the literature, a linear non-minimal coupling can give rise to successful inflation and efficient heating of the Universe in the case of a quadratic inflationary potential. Additionally, the gravitationally induced decay of the inflaton field can also source the whole observed DM abundance and baryon asymmetry of the Universe within simple particle physics frameworks.

%%%%%%%%%%%%%%%%%%%%%%%%%%%%%%%%%%%%%%%%%%%
\acknowledgments
%%%%%%%%%%%%%%%%%%%%%%%%%%%%%%%%%%%%%%%%%%%
BB would like to acknowledge `$\mathcal{C}$osmo $\mathcal{B}$eer' (IFT-UW) members for all their help and support through thick and thin.  
NB received funding from the Spanish FEDER / MCIU-AEI under the grant FPA2017-84543-P. JR is supported by a Ramón y Cajal contract of the Spanish Ministry of Science and Innovation with Ref. RYC2020-028870-I. This work was supported by the project PID2022-139841NB-I00 of MICIU/AEI/10.13039/501100011033 and FEDER, UE.

%%%%%%%%%%%%%%%
\appendix
%%%%%%%%%%%%%%%%

%%%%%%%%%%%%%%%%%%%%%%%%%%%%
\section{Inflaton Interactions and Decays}
\label{sec:decay-branching}
%%%%%%%%%%%%%%%%%%%%%%%%%%%%
In order to compute the gravity-induced inflaton decays, it is important to couple the inflaton to SM (and possible NP) fields, and therefore, the total action reads
\begin{equation}
    S = S_\phi + S_\text{SM} + S_\text{np}\,,
\end{equation} 
where $S_\phi$ was defined in Eq.~\eqref{eq:Sfinal}, the SM part is
\begin{align}
    S_\text{SM}= & \int d^4x\, \sqrt{-g}\, \Bigg[-\frac14\, g^{\mu\nu}\, g^{\lambda\rho}\, \mathcal{V}_{\mu\lambda}^{(a)}\, \mathcal{V}^{(a)}_{\nu\rho} + \frac{1}{F^2} \left(\mathcal{L}_Y - V(H)\right) \nonumber\\
    &\hspace{2.2cm} + \frac{1}{F}\, \big|D_\mu H\big|^2 + \frac{i}{F^{3/2}}\, \overline{f}\, \slashed{\partial}\, f + \frac{3\,i}{F^2}\, \overline{f} \left(\slashed{\partial}\, \Omega\right) f\Bigg],
\end{align}
where $\mathcal{V}^{(a)}$ denotes the SM gauge bosons (Abelian and non-Abelian) and $f$ stands for all SM fermions (quarks and leptons). The covariant derivative is defined as $D_\mu \equiv \partial_\mu - i\, g_2\, \tau^a\, W_\mu^a - i\, (g_1/2)\, Y\, B_\mu$, where $W_\mu$ and $B_\mu$ are the $SU(2)_L$ and $U(1)_Y$ gauge bosons, respectively, with corresponding $g_2$ and $g_1$ gauge coupling strengths, $Y$ is the hypercharge and $\tau^a = \sigma^a/2$ are the Pauli matrices. The potential of the Higgs doublet $H$ reads $V(H) = -\mu_H^2\, |H|^2 + \lambda_H\, |H|^4$. Additionally, the new physics sector is encoded in $S_\text{np}$, and may consist of a singlet scalar $S$ with mass $m_S$, a singlet Majorana neutrino $N$ with mass $m_N$, a Dirac fermion $\psi$ with mass $m_\psi$, or an Abelian gauge boson $X_\mu$ with mass $m_X$. In each case, the corresponding action in the Einstein frame reads
\begin{alignat}{2}
    & S_{S} = \int d^4x\,\sqrt{-g}\,\left[\frac{1}{2\,F}\,g^{\mu\nu}\,\partial_\mu S\,\partial_\nu S -\frac{1}{F^2}\,V(S)\right] &&\text{for scalar}, \\
    & S_N = \int d^4x\,\sqrt{-g} \left[\frac{i}{F^{3/2}}\, \overline N\, \gamma^\mu \partial_\mu N - \frac{1}{2\,F^2}\, m_N\, \overline{N^c}\, N - \left(\frac{y_N}{F^2}\, \overline{N}\, \widetilde H^\dagger\, L + \text{H.c}\right)\right] && \text{for Majorana},\\
    & S_{\psi} = \int d^4x\,\sqrt{-g}\,\left[\frac{i}{F^{3/2}}\,\bar\psi\,\gamma^\mu\partial_\mu\psi-\frac{1}{F^2}\,m_\psi\,\bar\psi\,\psi\right] && \text{for Dirac}, \\
    & S_X = \int d^4x\,\sqrt{-g}\,\left[-\frac{1}{4}\,X_{\mu\nu}\,X^{\mu\nu}+\frac{m_X^2}{2\,F^2}\,X_\mu\,X^\mu\right] &\quad& \text{for vector}.
\end{alignat}
For the scalar $S$ we disregard possible trilinear and quartic self-interactions, and also the mixing to the SM Higgs boson. In addition, for the vector $X^\mu$ we ignore its kinetic mixing with the SM $U(1)_Y$ gauge boson. Finally, if the new state is associated with DM, a $\mathbb{Z}_2$ parity is imposed under which only DM is odd to make it stable.
All relevant vertices were computed using {\tt LanHEP}~\cite{Semenov:2014rea} and are summarized in Table~\ref{tab:vertex}. The corresponding partial decay widths computed using {\tt CalcHEP}~\cite{Belyaev:2012qa} are reported in Eq.~\eqref{eq:Gamma}.
%%%%%%%%%%%%%%%%%%%%%%%%%%%%%%%%%%%%%%%%%%
\begin{table}[t!]
    \centering
    \begin{tabular}{|c|c|}
        \hline
        Interaction & Vertex\\
        \hline\hline
        $\phi\,\varphi\,\varphi$ & $\frac{\xi}{M_P} \left(p_i \times p_j + 2\, m_\varphi^2\right)$ \\
        \hline
        $\phi\,\Psi\,\Psi$ & $\frac{\xi}{2\,M_P} \left(4\,m_\Psi-3\,\slashed{p}_i\right)$ \\
        \hline
        $\phi\,V^\mu\,V^\nu$ & $\frac{-2\, \xi}{M_P}\, m_V^2\,\eta^{\mu\nu}$ \\
        \hline
    \end{tabular}
    \caption{Vertices for the inflaton-matter interactions, for scalars ($\varphi$), fermions ($\Psi$), and vectors ($V$).}
    \label{tab:vertex}
\end{table}
%%%%%%%%%%%%%%%%%%%%%%%%%%%%%%%%%%%%%%%%%%%%%

Note that, in order to highlight the impact of gravitationally induced decays, we have explicitly disregarded the existence of possible direct couplings among the inflaton and matter sectors.  However, the present analysis remains valid as long as the induced decay channels dominate over the direct ones.
For example, if the assumed linear $\mathbb{Z}_2$ symmetry of the inflaton sector is relaxed to include a direct inflaton-scalar coupling $\mu\, \tilde\phi\, \varphi^2$ leading to an Einstein-frame contribution $\mu\, \phi\, \left(1 + \xi\phi/(4\, M_P)\right)\varphi^2$, the ratio of direct to nonminimal gravitational production through decays becomes 
 \begin{equation}
     \frac{\Gp^\text{direct}}{\Gp^\text{non-minimal}} \sim \frac{\mu^2\, M_P^2}{\xi^2\, m_\phi^4}\,.
 \end{equation}
The non-minimally gravitational induced decay thus dominates if
\begin{equation} \label{eq:boundmu}
    \mu < \frac{m_\phi^2\, \xi}{M_P} \lesssim 1.6\times 10^{7}~\text{GeV} \left(\frac{\xi}{10^{-1}}\right)^2, 
\end{equation}
where we have taken into account Eq.~\eqref{eq:mphi-N} for the mass of inflaton with $N=50$ $e$-folds and the lower bound on $\xi$ from reheating (cf. Fig.~\ref{fig:bound}).
This bound was implicitly used for the case of Higgs production and scalar DM.

Furthermore, for the case of the Higgs boson, direct inflaton-Higgs trilinear ($\mu\, \phi\, H^\dagger H$) and quartic ($\lambda\, \phi^2\, H^\dagger H$) couplings could destabilize the required flatness of the inflationary potential. To ensure the stability of the Higgs vacuum during both inflation and preheating, it is required that $\mu \lesssim 10^8$~GeV and $\lambda \lesssim 10^{-8}$~\cite{Enqvist:2016mqj, Yang:2023rnh}. 

%%%%%%%%%%%%%%%%%%%%%%%%%%%%
\section{CP Asymmetry}
\label{sec:cp}
%%%%%%%%%%%%%%%%%%%%%%%%
The CP asymmetry generated from $N_1$ decay is given by~\cite{Davidson:2008bu}
\begin{equation}\label{eq:cp1}
    \epsilon_{\Delta L} \equiv \frac{\Gamma_{N_1 \to \ell_i\, H } -\Gamma_{N_1 \to \bar\ell_i\, \bar H}}{\Gamma_{N_1 \to \ell_i\, H} + \Gamma_{N_1 \to \bar\ell_i\, \bar H}} \simeq \frac{1}{8\, \pi}\, \frac{1}{(y_N^\dagger\, y_N)_{11}}\, \sum_{j=2, 3} \text{Im}\left(y_N^\dagger\, y_N\right)^2_{1j} \times \mathcal{F}\left(\frac{M_j^2}{M_1^2}\right),
\end{equation}
where
\begin{equation}
    \mathcal{F}(x) \equiv \sqrt{x}\,\left[\frac{1}{1-x}+1-(1+x)\,\log\left(\frac{1+x}{x}\right)\right].
\end{equation}
For $x\gg 1\,,\mathcal{F}\simeq -3/\left(2\,\sqrt{x}\right)$, and Eq.~\eqref{eq:cp1} becomes
\begin{equation}
    \epsilon_{\Delta L} \simeq -\frac{3}{16\, \pi}\, \frac{1}{(y_N^\dagger\, y_N)_{11}} \left[\text{Im}\left(y_N^\dagger\, y_N\right)^2_{12} \frac{m_{N_1}}{m_{N_2}} + \text{Im}\left(y_N^\dagger\, y_N\right)^2_{13} \frac{m_{N_1}}{m_{N_3}}\right].
\end{equation}
If we consider $\text{Im}\left(y_N^\dagger\, y_N\right)^2_{13} \gg \text{Im}\left(y_N^\dagger\, y_N\right)^2_{12}$ and $m_{N_1}\ll m_{N_{2,3}}$, then 
\begin{equation}
    \epsilon_{\Delta L} \simeq -\frac{3\, \delta_\text{eff}}{16\, \pi}\,\frac{|(y_N)_{13}|^2 \, m_{N_1}}{m_{N_3}}\,,   
\end{equation}
while the effective CP violating phase is given by
\begin{equation}
    \delta_\text{eff} = \frac{1}{(y_N)_{13}^2}\, \frac{\text{Im}(y_N^\dagger\,y_N)^2_{13}}{(y_N^\dagger\,y_N)_{11}}\,.    
\end{equation}
In order to simultaneously generate the tiny active neutrino mass, one has to impose the seesaw relation
\begin{equation}
    m_{\nu_3} = \frac{|(y_N)_{13}|^2\, v^2}{m_{N_3}}\,,    
\end{equation}
that leads to
\begin{equation}
    \epsilon_{\Delta L} \simeq -\frac{3\, \delta_\text{eff}}{16\, \pi}\, \frac{m_{N_1}\, m_{\nu_3}}{v^2}\,.
\end{equation}
Instead, if $\text{Im}\left(y_N^\dagger\, y_N\right)^2_{13} \ll \text{Im}\left(y_N^\dagger\, y_N\right)^2_{12}$, the CP asymmetry
parameter becomes
\begin{equation}
    \epsilon_{\Delta L} \simeq -\frac{3\, \delta_\text{eff}}{16\, \pi}\, \frac{m_{N_1}\, m_{\nu_2}}{v^2}\,.    
\end{equation}
In general, one can then write
\begin{equation}
    \epsilon_{\Delta L} \simeq -\frac{3\, \delta_\text{eff}}{16\, \pi}\, \frac{m_{N_1}\, m_{\nu_i}}{v^2}\,,    
\end{equation}
where $i=2\,,3$ for normal hierarchy. On a similar note, the CP-asymmetry parameter can be obtained for the inverted hierarchy with $i=1\,,2$. In either case, we consider $m_{\nu_i}$ to be the heaviest active neutrino mass $m_{\nu,{\rm max}}$ in Eq.~\eqref{eq:cp}.

% %%%%%%%%%%%%%%%%%%%%%%%
% \section{Additional Terms}
% %%%%%%%%%%%%%%%%%%%%%%%%
% In presence of scalar singlet DM, direct inflaton-scalar coupling can arise without violating the $\mathbb{Z}_2$ symmetry: $\mu_{\rm dir}\,\tilde\phi\,\varphi^2\to\mu_{\rm dir}\,\phi\,\left(1+\xi\phi/(4\,M_P)\right)\varphi^2$. We compare the DM production rate via direct decay to decay through non-minimal coupling as
% \begin{align}
% & \frac{\Gamma_{\rm dir}}{\Gamma_{\rm nm}}\simeq\frac{\mu_{\rm dir}^2\,M_P^2}{\xi^2\,m_\phi^4}\,,
% \end{align}
% where $\Gamma_{\rm nm}$ is the decay width arising due to non-minimal coupling and $\Gamma_{\rm dir}$ is that due to direct coupling. The gravitationally induced decay thus dominates if
% \begin{align}
% & \mu_{\rm dir}<\frac{m_\phi^2\,\xi}{M_P}\implies \frac{\mu_{\rm dir}}{\text{GeV}}\lesssim 1.6\times 10^{7}\,\left(\frac{\xi}{10^{-1}}\right)^2\,, 
% \end{align}
% where we have used the lower bound on $\xi$ from reheating [cf. Fig.~\ref{fig:bound}] and Eq.~\eqref{eq:mphi-N} for the mass of inflaton with $N=50$. 

% One can similarly have direct inflaton-Higgs trilinear coupling $\mu_{\phi h}\,\phi\,H^\dagger H$ or quartic coupling $\lambda\,\phi^2\,H^\dagger H$. It has been shown in~\cite{Enqvist:2016mqj,PhysRevD.108.103524}, for quadratic potential for the inflaton during reheating, $\mu_{\phi h}\lesssim 10^8$ GeV and $\lambda< 10^{-8}$, ensuring stability of the Higgs vacuum both during inflation and during preheating. 

%%%%%%%%%%%%%%%%%%%%%%%%%
\bibliographystyle{JHEP}
\bibliography{biblio}
%%%%%%%%%%%%%%%%%%%%%%%%%
\end{document}